\documentclass{article}
\usepackage{icrctc07}

\title{Measurement of the UHECR energy spectrum from hybrid
 data of the Pierre Auger Observatory}
\shorttitle{UHECR energy spectrum from hybrid
 data of the Pierre Auger Observatory}

\authors{Lorenzo Perrone~$^1$, for the the Pierre Auger Collaboration~$^2$}
\shortauthors{Author and et al.}
\afiliations{$^1$ Universit\`a del Salento and INFN Lecce, I-73100 Lecce, Italy\\
$^2$ Observatorio Pierre Auger, Av. San Mart\'in Norte 304, Malarg\"ue, (5613) Mendoza, Argentina}
\email{lorenzo.perrone@le.infn.it}

\abstract{More than two years of fluorescence detector data collected in coincidence with at
least one station of the surface detector array (``hybrid data'') are used
to measure the flux and energy spectrum of cosmic rays above about 10$^{18}$
eV. The hybrid measurement extends towards lower energies the spectrum
measured with the surface detector data only, and provides a cross-check
with an independent data set. The determination of the fluorescence
detector aperture and of its live-time, which is the major aspect of this
measurement, is illustrated in detail. Our current estimate of the
corresponding systematic uncertainties are given.}

\begin{document}

\maketitle

\section{Introduction}

The Pierre Auger Observatory employs two independent detection
techniques, allowing the reconstruction of extensive air showers with
two complementary measurements. Indeed, the combination of information from
the surface array and the fluorescence telescopes enhances the
reconstruction capability of ``hybrid'' events with respect to the individual
detector components.
A description of the hybrid performance of the Pierre Auger Observatory
is given in~\cite{dawson}.\\
In this analysis, the energy spectrum of cosmic rays is measured using 
 hybrid data collected between December 2004 and February 2007. 
 The inspected energy range covers a region where the transition from Galactic to extra-galactic  
 cosmic rays is expected to occur.\\  
Due to construction, the configuration of fluorescence telescopes and surface detector 
has evolved significantly and the effective detection area has correspondingly changed. 
The key points of  
the analysis are an accurate estimate of the hybrid detector exposure and an appropriate    
selection of well-reconstructed events. A good knowledge of systematic uncertainties 
is also required to support the robustness of the results.  

\section{Hybrid Exposure}
\begin{figure}[t]
  \begin{center}
    \includegraphics[width=0.43\textwidth]{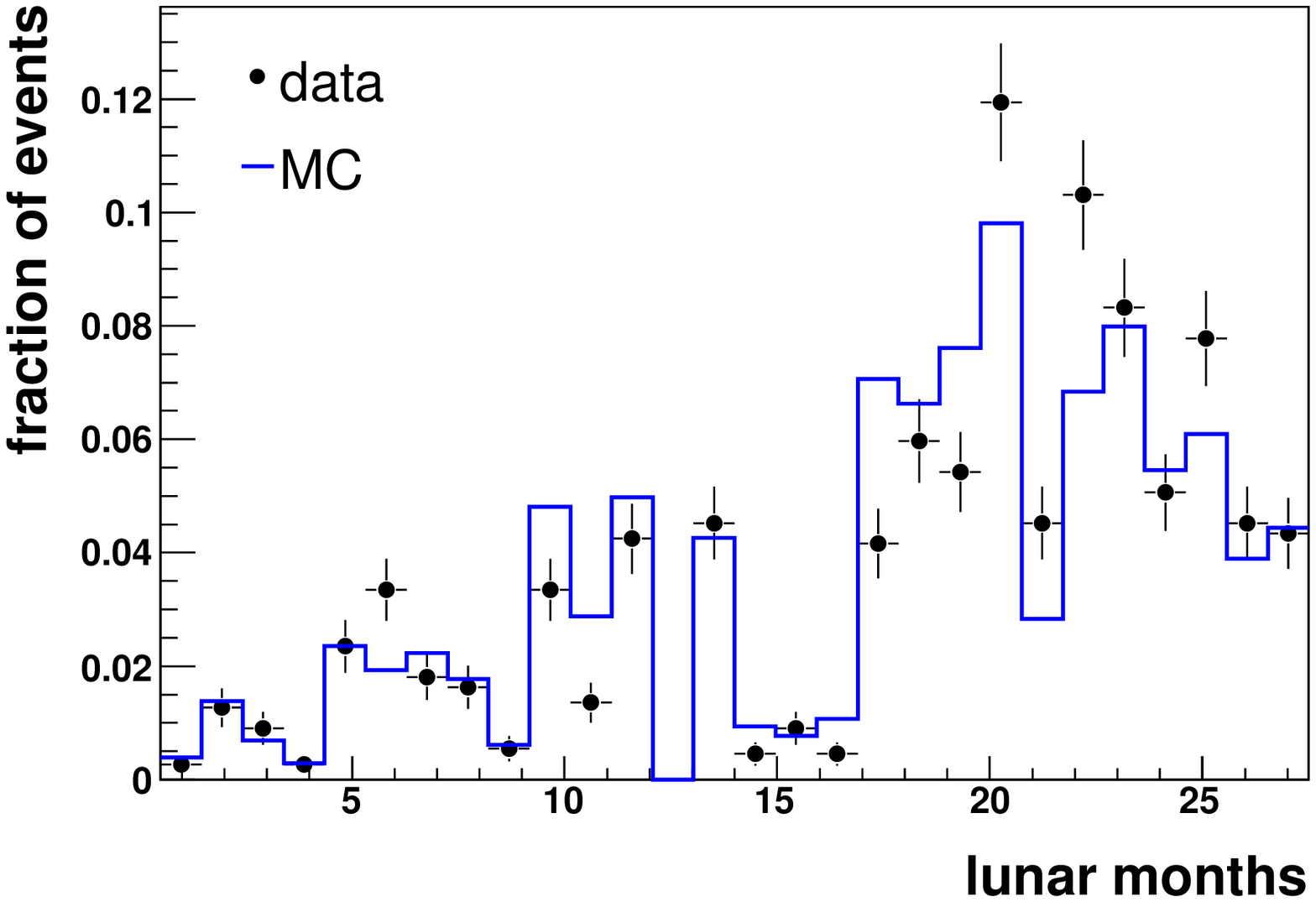}
    \includegraphics[width=0.43\textwidth]{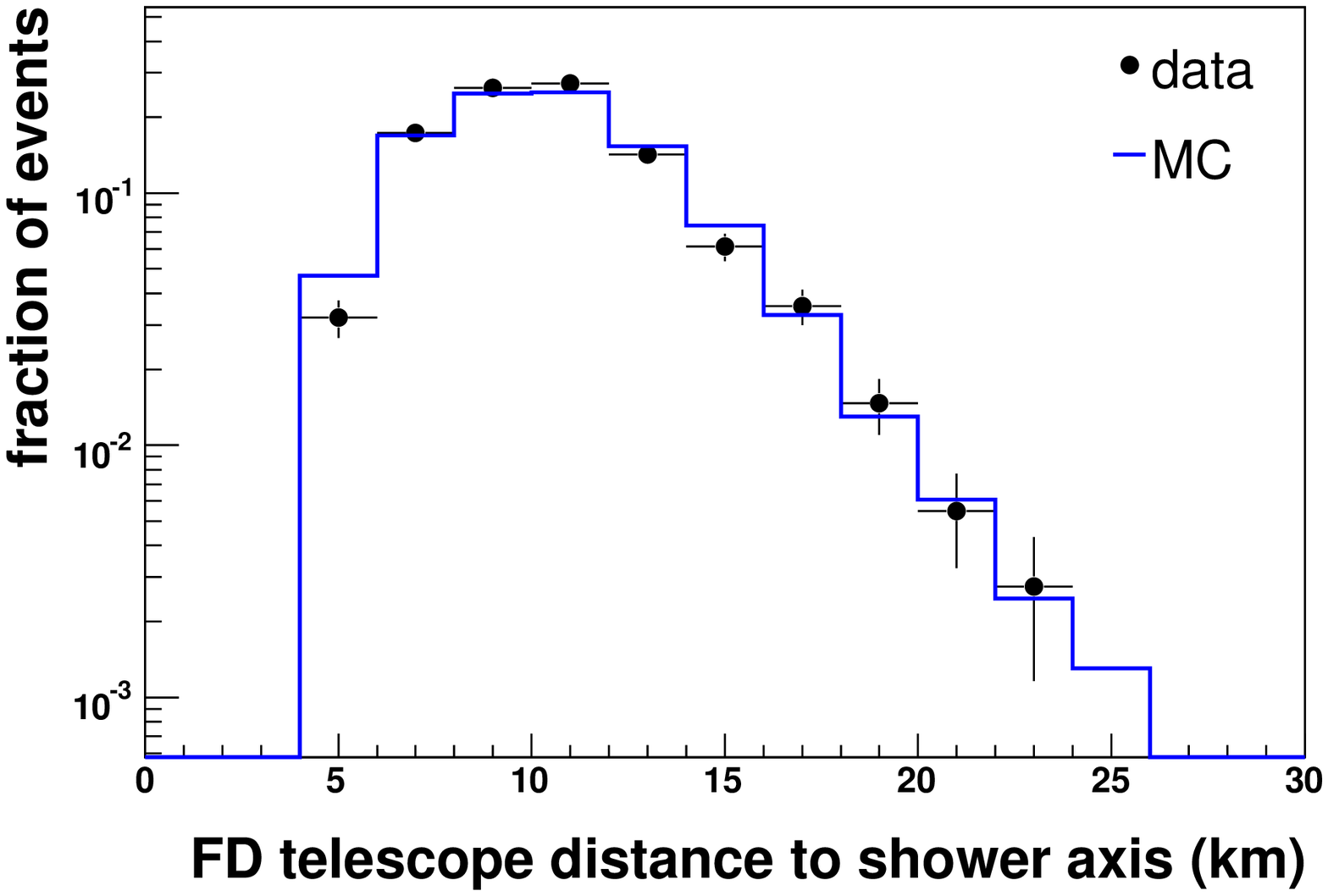}
    \caption{Fraction of events as a function of lunar months 
     (top) 
     and distribution of telescope distance to shower axis (bottom),    
     for data and simulation (same selection cuts applied).}    
    \label{fig:sim-data}
  \end{center}
\end{figure}

\begin{figure}[t]  
  \begin{center}
  \vskip -0.3 cm
    \includegraphics[width=0.44\textwidth]{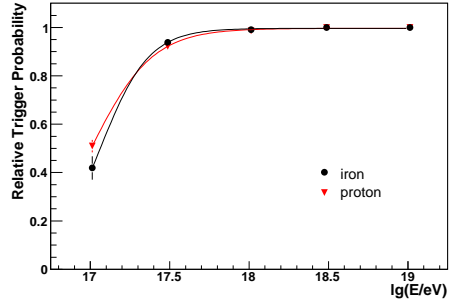}
    \caption{Hybrid trigger efficiency for proton and iron 
  (full simulation method)}    
    \label{fig:fullhyb}
  \end{center}
\end{figure}

The calculation of the hybrid exposure
relies on a detailed simulation of fluorescence (FD) 
and surface detector (SD) response. 
To reproduce 
the exact working conditions of the experiment 
and the entire sequence of given configurations, a large sample of 
Monte Carlo simulations have been performed.
Several factors (fast 
growth of surface array and ongoing extension of 
the fluorescence detector, seasonal and instrumental effects) can introduce 
a significant dependence 
of aperture on time. This effect has been     
taken into account and simulated using an accurate calculation 
of the hybrid detector uptime.   
The simulation sample consists of 
a large number of longitudinal energy deposit profiles generated  with 
CONEX~\cite{conex}.       
The energy spectrum ranges from $10^{17}$eV  to $10^{21}$eV according
  to a power-law function with differential spectral index -2 (reweighted to -2.8 
  when comparing data to simulation)
and the zenith angles are sampled between 0$^\circ$ and 70$^\circ$.
Fig.~\ref{fig:sim-data} (top) shows the number of collected 
events as a function of lunar months (FD measurement cycles after December
2004) for data and simulation. There is a good overall agreement along the entire time scale 
  considered for this analysis.     
The simulation has been validated by comparing the distribution of reconstructed  
observables to experimental data.   
Fig.~\ref{fig:sim-data} (bottom) shows the distribution of the 
telescope distance to shower axis,  
for data and simulation. A very good agreement is found at this selection level. \\ 
The distribution of particles at ground is not provided by CONEX.  
Nevertheless, the time of the station 
with the highest signal is sufficient information for this analysis. 
This time is used in 
the hybrid reconstruction for determining the 
incoming direction of the showers, and the impact point at ground.  
Once the shower geometry is known, 
the longitudinal profile can be reconstructed and the energy calculated.
The tank trigger simulation is performed using a parameterisation  
based on ``Lateral Trigger Probability'' functions (LTPs)~\cite{ltpicrc2005}. 
They give the probability for a shower to trigger a tank as a
 function of primary cosmic ray energy, mass, direction and tank 
distance to shower axis.
A full hybrid simulation with CORSIKA showers~\cite{corsika} 
(FD and SD response are simultaneously and fully simulated) 
has shown that the hybrid trigger efficiency 
(a fluorescence event in coincidence with at least one tank)    
is flat and equal to 1 at energies greater than $10^{18}$ eV. 
This feature is shown in Fig.~\ref{fig:fullhyb}  
 for proton and iron primaries. 
For these energies, the hybrid trigger efficiency   
 coincides with the one derived from the LTPs based method.
The difference between the two primaries 
becomes negligible at energy larger than $10^{17.5}$ eV.
A detailed description of the hybrid detector 
simulation program is given 
in~\cite{offline}.
\section{Data Selection}
\begin{figure}[h!]
  \begin{center}
    \includegraphics[width=0.46\textwidth]{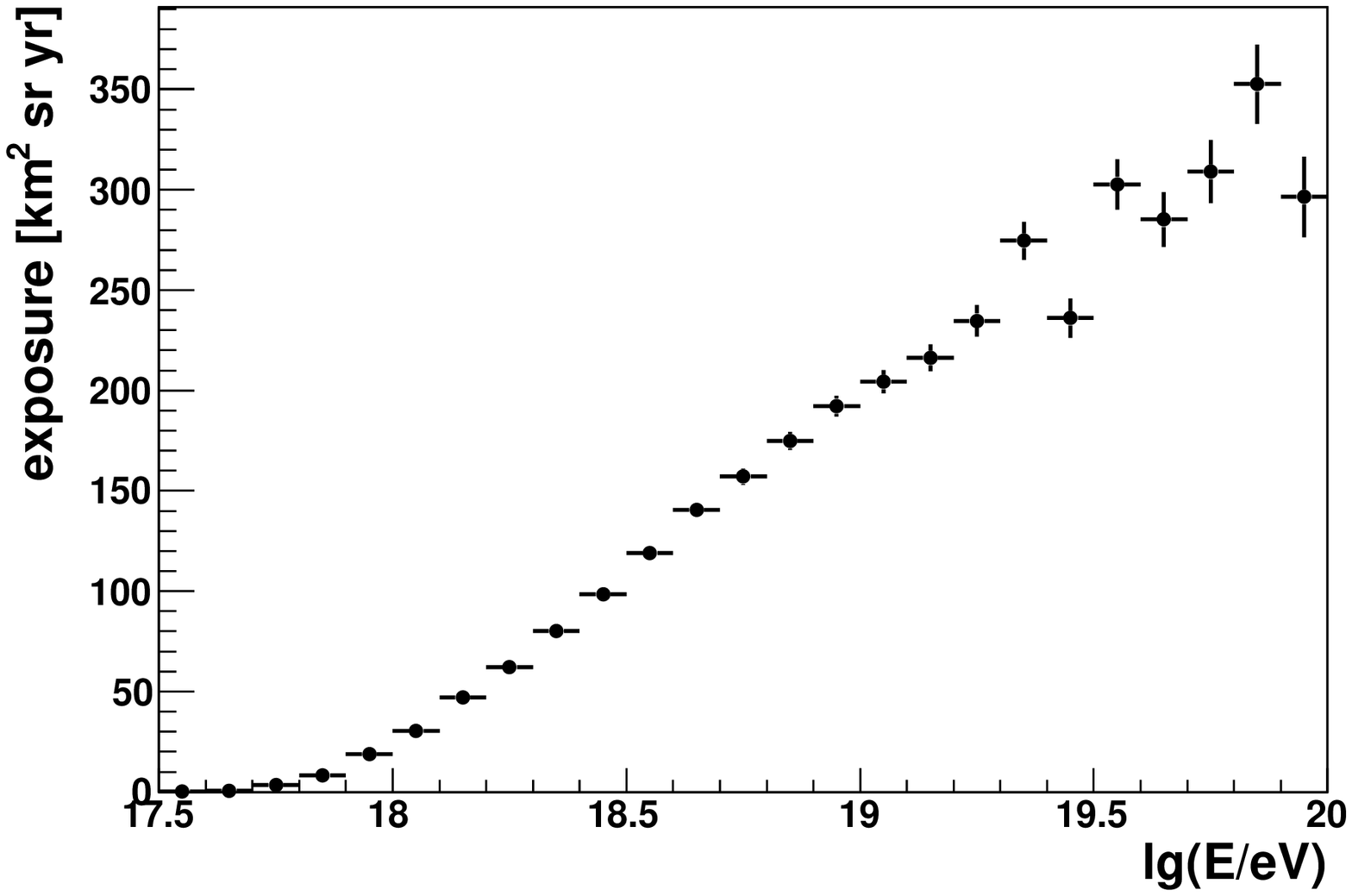}
    \includegraphics[width=0.425\textwidth]{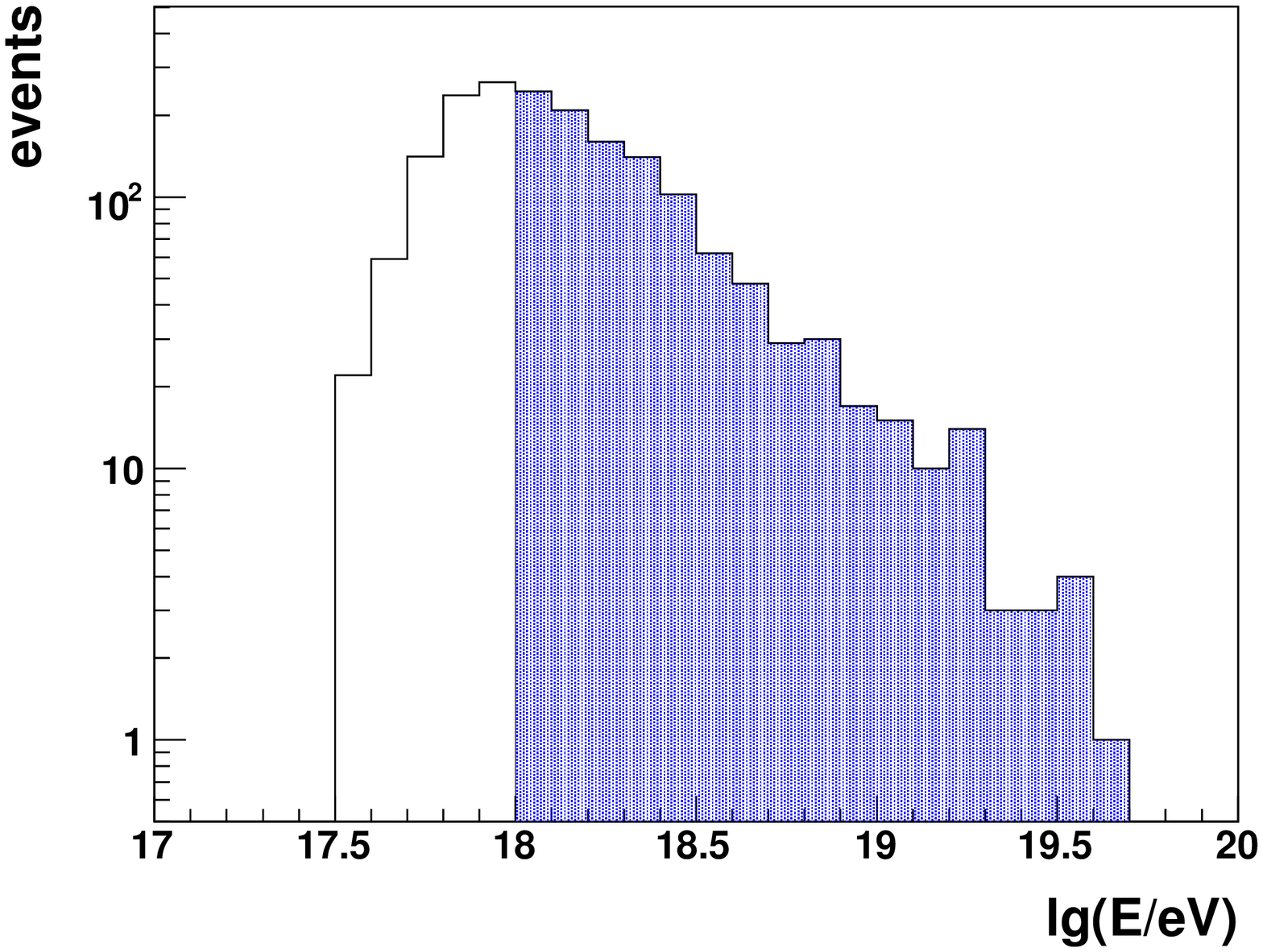}
    \caption{Hybrid exposure after all cuts (top).
     Energy distribution of selected data (bottom). 
       The number
      of events used for the spectrum (E $>10^{18}$ eV, shadowed area) is 1092.}
    \label{fig:exp}
  \end{center}
\end{figure}

\begin{figure*}[t]
  \begin{center}
    \includegraphics[width=0.48\textwidth]{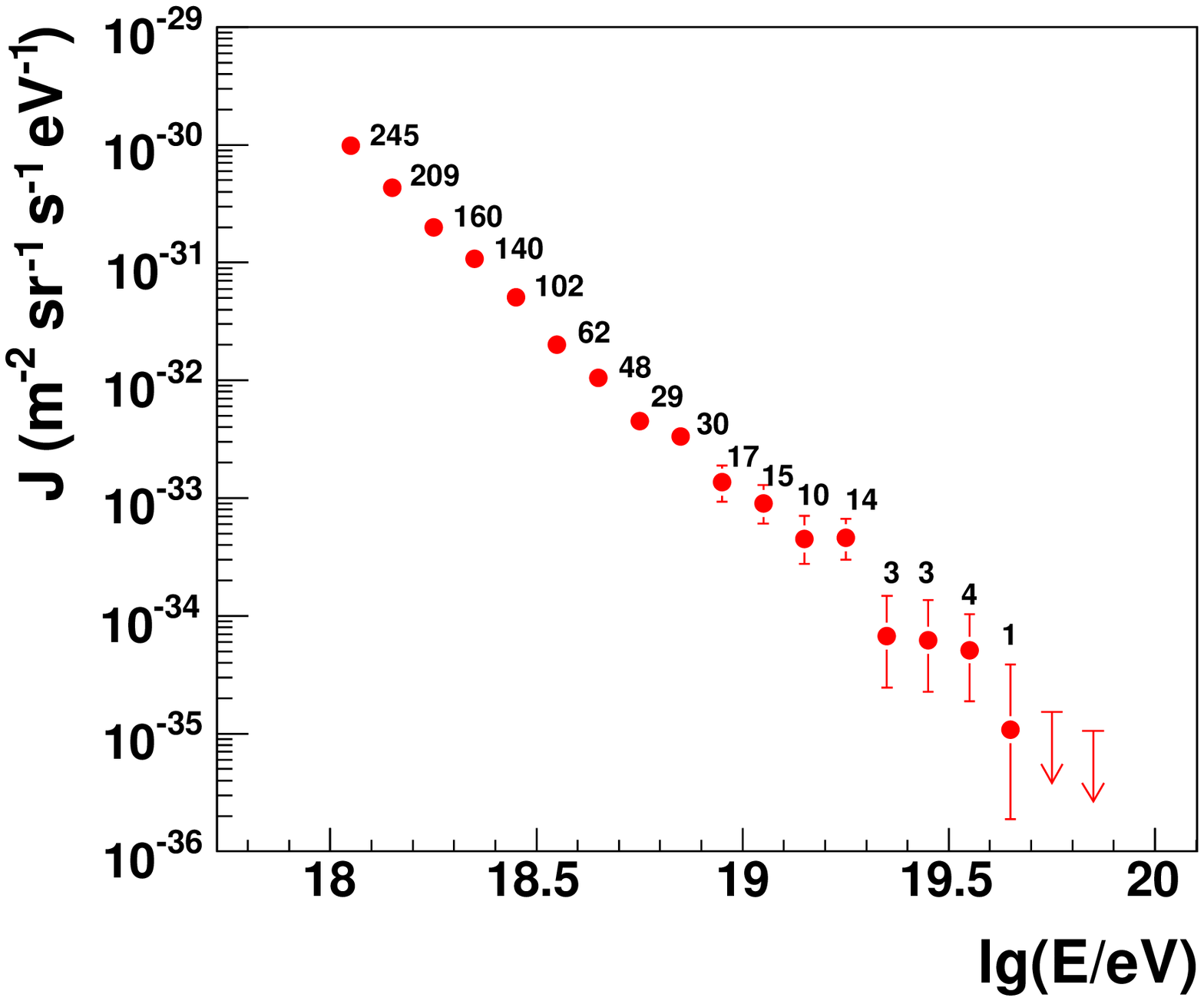}
    \includegraphics[width=0.47\textwidth]{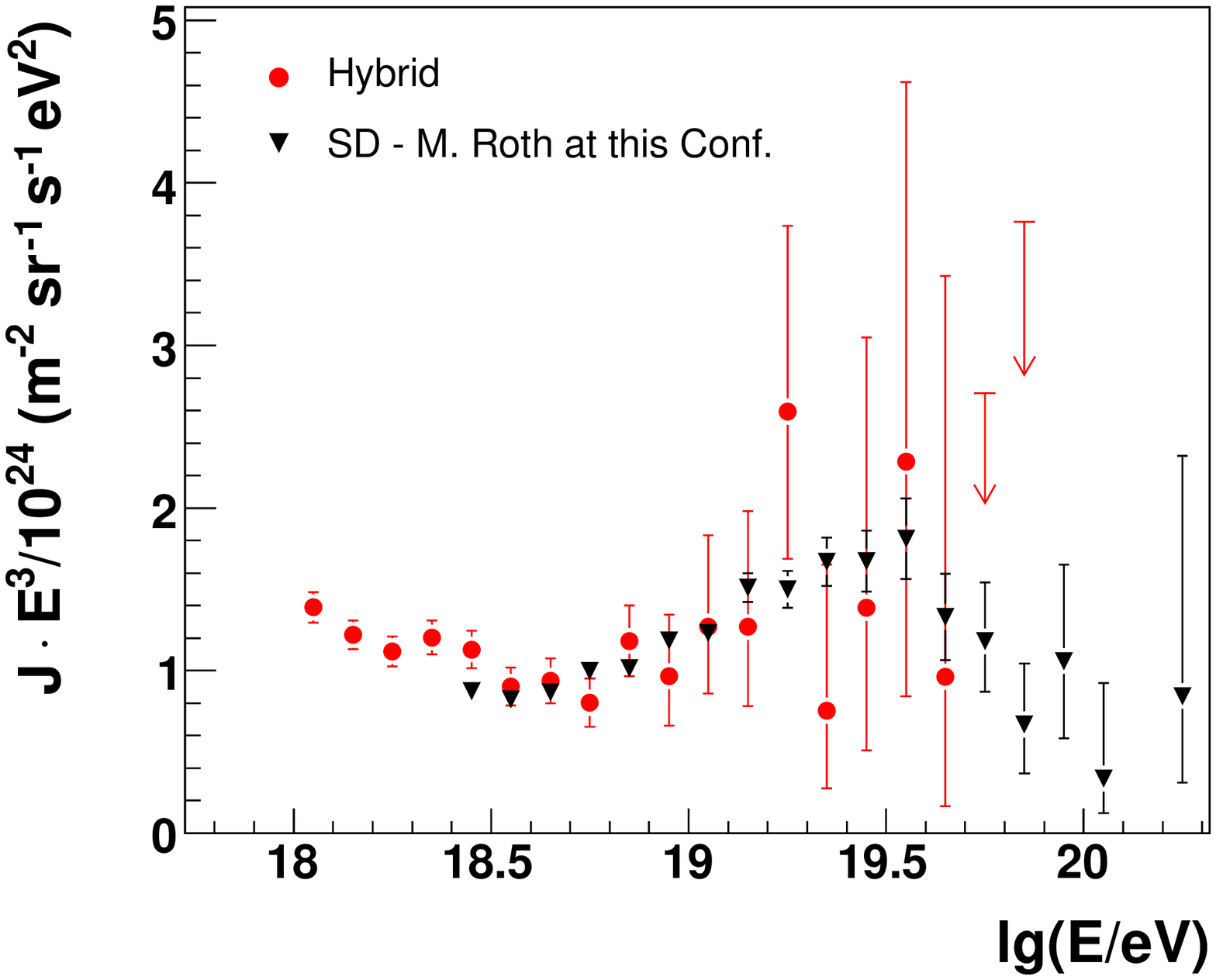}
    \caption{Hybrid energy spectrum (left) shown in comparison (right) with 
     surface detector spectrum (only statistical uncertainties are given in the figure).}
    \label{fig:spec}
  \end{center}
\end{figure*}

Only data with a successful hybrid geometry reconstruction are selected 
for calculating the hybrid spectrum.  
To suppress monocular events with random surface detector triggers,  
 only events with the station used for reconstruction lying within 750~m 
 from the shower axis are accepted. 
 This condition ensures that the probability of the station to trigger is equal to one. 
 Showers that are expected to develop outside the geometrical 
 field of view of the 
 fluorescence detectors are also rejected and, based on data, a fiducial volume for detection  
  is defined as a function of the reconstructed energy. Details on how the fiducial 
  volume is taken are given in~\cite{ER} and \cite{photlim}.
   Moreover, 
  only events with reconstructed zenith angle less than 60$^{\circ}$ are accepted.
The observed profile and reconstructed shower depth at maximum ($X_{max}$)
are required to satisfy the following conditions:\\
- a successful Gaisser-Hillas fit with $\chi^{2}$/Ndof $<$ 2.5 for the
 reconstructed longitudinal profile\\
- minimum observed depth $<$ $X_{max}$ $<$ maximum observed depth\\
- a relative amount of Cherenkov light in the signal less than 50\%\\
- measurement of atmospheric parameters available.\\
A fluorescence photon yield according to~\cite{nagano} is currently used for 
energy reconstruction.
Finally, as the algorithm used for the profile reconstruction
 propagates both, light flux and geometrical uncertainties,  
the estimated uncertainties of shower energy is a good variable to reject
poorly reconstructed showers. We require $\sigma(E)/E<20$\%. 
Fig.~\ref{fig:exp} shows the hybrid exposure (top) and the energy distribution of 
 all events (bottom) at the last reconstruction level (all quality cuts have been applied). 
Exposure at this level depends very weakly on chemical composition, giving 
a spectrum basically independent of any assumption on primaries mass.     
The hybrid spectrum deriving from this analysis is shown in Fig.~\ref{fig:spec} (left),   
compared (right) with the spectrum from surface detector presented 
in~\cite{priv} (only statistical 
 uncertainties are given in the figure).  

\section{Systematics}
\begin{figure}[h]
  \begin{center}
    \includegraphics[width=0.49\textwidth]{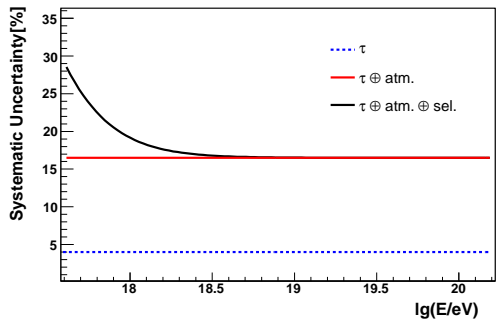}
    \caption{Systematic uncertainties on hybrid spectrum 
    due to live time ($\tau$), 
    atmospheric conditions (atm) and 
    impact of the energy scale uncertainty on events selection (sel).}
    \label{fig:sys}
  \end{center}
\end{figure}

The hybrid spectrum is primarily affected by the systematic uncertainty 
on the energy determination (about 22\%~\cite{dawson}). 
Further systematic uncertainties and their individual contributions 
are shown in Fig.~\ref{fig:sys} as a function of energy. 
The calculation of detector uptime has been independently  
cross-checked using the observed laser shots fired by the Central Laser Facility (CLF)~\cite{clf} 
and the results agree 
at the level of 4\%.
A more significant source of uncertainty (16 \%) is expected from the lack of  
a precise knowledge of atmospheric conditions.  
Part of the shower profile 
may be shadowed by clouds or the Cherenkov light can be diffused by 
fog and/or clouds and redirected towards the detector. This uncertainty is still 
large but it is expected to be significantly reduced when all atmospheric 
monitoring data have been fully analysed.
Finally, an uncertainty, increasing at lower energies, 
is expected 
as a consequence 
of the aperture calculation at reconstruction level. Indeed, at low energy, 
the efficiency of the event selection algorithm 
varies rapidly with energy and is very sensitive to a systematic energy shift. 
An overall uncertainty (all contribution summed up in quadrature) 
of about 20\% is expected at E=10$^{18}$ eV (see Fig.~\ref{fig:sys}). 
As a final remark,
it is worth saying that the extension to the viewing elevations of FD telescopes  
will allow to be reached lower energies with smaller systematics~\cite{HEAT}.

\section{Conclusions}
More than two years of hybrid data (fluorescence events 
    in coincidence with at least one station) have been used
to measure the energy spectrum of cosmic rays above 10$^{18}$
eV. Very good agreement with the spectrum
measured by the surface detector is found within the estimated 
FD systematic uncertainties. A combined spectrum is presented in~\cite{toko} and astrophysical implications 
are also discussed there.

\end{document}